\documentclass{article}
\usepackage{spconf,amsmath,graphicx}

\usepackage{enumitem}
\setlist{nosep, leftmargin=14pt}

\usepackage{mwe} 

\usepackage{stfloats}

\usepackage[colorlinks,
            linkcolor=blue,
            anchorcolor=blue,
            citecolor=blue]{hyperref}

\title{CSDN: Combining Shallow and Deep Networks for Accurate Real-time Segmentation of High-definition Intravascular Ultrasound Images}
\name{Shaofeng Yuan$^{1 *}$ \qquad Feng Yang$^{2 *}$ \thanks{* Shaofeng Yuan and Feng Yang are the corresponding authors (shaofeng.yuan.smu@gmail.com; yangf@smu.edu.cn).}}
\address{$^{1\ }$Institute of Artificial Intelligence, Insight Lifetech, Shenzhen, China \\
         $^{2\ }$School of Biomedical Engineering, Southern Medical University, Guangzhou, China}
\begin{document}
%
\maketitle
\begin{abstract}
Intravascular ultrasound (IVUS) is the preferred modality for capturing real-time and high resolution cross-sectional images of the coronary arteries, and evaluating the stenosis. Accurate and real-time segmentation of IVUS images involves the delineation of lumen and external elastic membrane borders. In this paper, we propose a two-stream framework for efficient segmentation of 60 MHz high resolution IVUS images. It combines shallow and deep networks, namely, CSDN. The shallow network with thick channels focuses to extract low-level details. The deep network with thin channels takes charge of learning high-level semantics. Treating the above information separately enables learning a model to achieve high accuracy and high efficiency for accurate real-time segmentation. To further improve the segmentation performance, mutual guided fusion module is used to enhance and fuse both different types of feature representation. The experimental results show that our CSDN accomplishes a good trade-off between analysis speed and segmentation accuracy.
\end{abstract}
\begin{keywords}
Shallow network, Deep network, Real-time segmentation, Intravascular ultrasound images, Medical image segmentation
\end{keywords}
\section{Introduction}
\label{sec:intro}
Atherosclerosis is a disease of the vessel wall, and responsible for many cardiovascular diseases.
Compared with the in vitro screening, the widespread application of the intravascular ultrasound (IVUS) imaging relies on its capability to visualize the inner structure of vessels in real-time to diagnose the arteriosclerotic disease of the coronary artery.
It can assess quantitative clinical measurements.
However, accurate delineation of the lumen and external elastic membrane (EEM) borders is essential for the assessment of plaque burden and stenosis degree.
The current clinical practice relies on manual annotation in the IVUS images, which is time-consuming and user-dependent.

In recent years, many deep learning methods based on convolutional networks (ConvNets) have been widely used in computer vision and image processing tasks due to their excellent capacity of automatic feature extraction \cite{he2016,dong2014}.
Considerable progress has been made in medical image computing community \cite{tang2020}.
Many ConvNets-based methods have been developed to segment lumen and/or EEM regions.
For example, Ji et al. \cite{ji2019} proposed IVUS-Net series based on U-shaped fully convolutional networks \cite{ronneberger2015}.
Different from simple encoding and decoding layers in U-net, IVUS-Net uses aggregated, multi-branch architectures in these layers.
Cao et al. \cite{cao2020} selected DeepLabv3+ \cite{chen2018} as the segmentation network in their work.
It combines the advantages of spatial pyramid pooling module and encoder-decoder structure.
Xia et al. \cite{xia2020} proposed a multi-scale feature aggregated U-net (MFAUNet) to extract two membrane borders simultaneously.
The feature aggregated module in skip connections utilizes the bi-directional convolutional long short-term memory unit to extract the context information from the spatial-temporal perspective.
Ziemer et al. \cite{ziemer2020} proposed a multi-frame ConvNet for lumen segmentation. Adding information about neighbouring frames surrounding the frame of interest improved the segmentation performance.
Inspied by this work, we adopt three-frame IVUS images as input of our CSDN.
Li et al. \cite{li2021} used two modified U-net for lumen and media-adventitia borders, respectively. The results of two segmentations are combined in the end.
Szarski et al. \cite{szarski2021} proposed a real-time modified U-net augmented with learned translation dependence (coordinate-aware FCN, CoordFCN).

However, the above methods don't accomplish a good trade-off between processing speed and segmentation accuracy. In this paper, a novel framework CSDN is proposed to timely and accurately extract two important borders in IVUS images by considering that low-level details and high-level semantics are crucial to the real-time segmentation task, and treated separately can achieve the trade-off between the accuracy and inference speed. 


\section{Methods}
\label{sec:methods}
\begin{figure*}[htb]
\begin{center}
\includegraphics[width=1.0\textwidth]{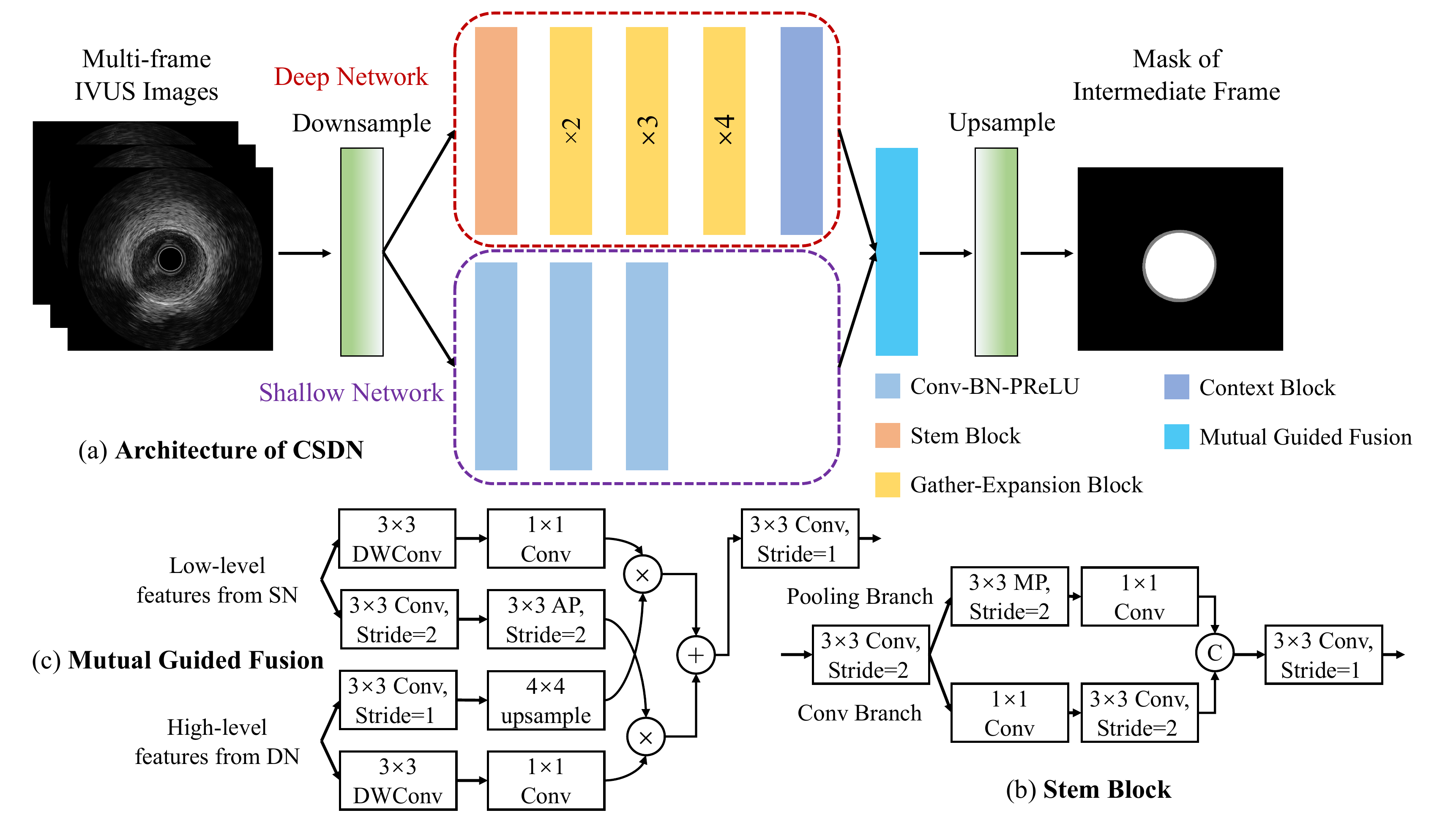}
\caption{(a) The architecture of the proposed CSDN. On the top side, Deep Network with thin channels is adopted to extract high-level semantic information. On the bottom side, Shallow Network with thick channels is used to extract low-level detailed information. At last, the output features from the above networks are input to Mutual Guided Fusion to generate final output feature and make the final prediction. Before feature extraction and label assignment, Downsample and Upsample modules are used to decrease and increase image size for accelerating segmentation. (b) Components of the Stem Block. (c) Components of the Mutual Guided Fusion. MP is max pooling, AP is average pooling. DWConv is depth-wise convolution. Batch normalization and PReLU are omitted.}
\label{fig1}
\end{center}
\end{figure*}

The overview of the proposed CSDN architecture is shown in Fig.~\ref{fig1} (a).

\subsection{Downsampling with pixel unshuffling}
\label{ssec:sub2.1}
Directly extracting visual features in high resolution space, ConvNets require more computational cost and memory footprint.
However, during percutaneous coronary intervention (PCI), accurate and real-time IVUS segmentation for lumen and EEM area is a requirement.
Shi et al. \cite{shi2016} proposed to use pixel shuffling as an upsampling operation, alternative to deconvolution layer.
ConvNets with pixel shuffling, first gets wide but low resolution feature maps, then arranges input channels to produce a feature map with higher resolution.
Contrary to pixel shuffling, pixel unshuffling is a downsampling operation, exchanging channel information with spatial information.
As shown in Fig.~\ref{fig1}, CSDN uses Downsample and Upsample modules to accelerate feature extraction of neural networks.
In practice, Downsample module include bicubic interpolation and pixel unshuffling.
After Downsample module, the width size of multi-frame IVUS images is reduced by a factor of 4.

\subsection{Shallow network}
\label{ssec:sub2.2}
As shown on the bottom side in Fig.~\ref{fig1}(a), the shallow network with thick channels focuses to extract low-level details.
This network has a shallow structure with thick channels in building blocks, because we enforce it to encode enough spatial detailed information.
The shallow network has three feature extraction blocks.
Each block has three standard 2d convolution layers.
Each convolution layer is followed by a 2d batch normalization and a PReLU activation function.
In each block, the first convolution layer has stride of 2 for downsampling the size of feature maps.
Obviously, the width size of output from the shallow network is reduced by a factor of 8.

\subsection{Deep network}
\label{ssec:sub2.3}
As shown on the top side in Fig.~\ref{fig1}(a), the deep network with thin channels takes charge of learning high-level semantics.
This network has a deep architecture with thin channels in building blocks for better segmentation performance.
In dense prediction task, e.g., semantic segmentation, sufficient receptive field is of importance for the good performance.
Therefor, we design deep network to provide sizeable receptive field using a fast down-sampling strategy.
The deep network has five context and semantic information feature extraction stages.
The first stage is the Stem Block, shown in Fig.~\ref{fig1}(b).
This block has two branches to downsample input feature maps in different manners.
In the end of two branches, both output feature maps are concatenated, then followed by a convolution layer with batch normalization and PReLU activation function.
The last stage is the Context Block \cite{yu2018,yu2021}.
This block uses residual connection and global average pooling to embed the global contextual information, providing the maximum receptive field.
The remaining three stages are Gather-Expansion Block \cite{yu2021}.
Each Gather-Expansion Block has at least 2 gather-expansion (GE) layers.
For example, the first Gather-Expansion Block has 2 GE layers, the second has 3, and the third has 4, shown in Fig.~\ref{fig1}(a) gold boxes.
The GE layer has 2 types, GE-Stride1 and GE-Stride2, details in \cite{yu2021}, and both are residual connections.
The first GE layer in each GE block uses GE-Stride2 for downsampling feature maps to enlarge receptive field.
The other GE layer in each GE block uses GE-Stride1.
In GE-Stride1, a 3 $\times$ 3 convolution is used to gather local feature values and expand to higher-dimensional space.
Then an efficient 3 $\times$ 3 depth-wise convolution is used independently over each channel from the above step.
In the end, a 1 $\times$ 1 convolution is used to project higher-dimensional feature maps into a low channel capacity space.

\subsection{Feature fusion with mutual guided fusion}
\label{ssec:sub2.4}
The feature representation of the shallow network and the deep network is different and complementary.
Thus, we use mutual guided fusion block to merge both types of feature representation, illustrated in Fig.~\ref{fig1}(c).
This block use contextual information from deep network to guide the feature value from shallow network.
Meanwhile, this block use detail information from shallow network to guide the feature response from deep network.
Comparing to the simple feature fusion like element-wise summation or feature concatenation, mutual guided fusion strategy enables efficient communication between both two neural networks.

\section{Experiments and Results}
\label{sec:experimentsandresults}

\subsection{Dataset}
\label{ssec:sub3.1}

In this work, we evaluate the proposed CSDN framework on High-definition Intravascular Ultrasound (HDIVUS) database, which is a partial and in-house dataset for IVUS image segmentation created by Insight Lifetech.
The dataset consists of 2098 sets of B-mode IVUS images, including a complete annotation of two important borders by expert cardiologists.
In total, the HDIVUS dataset contains IVUS images from 14 patients.
For each pullback, about 150 IVUS images were sampled.
All the original images have a size of 900 $\times$ 900 pixels.

\subsection{Evaluation metrics}
\label{ssec:sub3.2}
The evaluation metrics used in this work are two types, based on region and border.
The region-based metrics are Dice-Sørensen Coefficient (DSC), Intersection over Union(IoU) of EEM and lumen.
The border-based metrics is Hausdorff distance (HD) in the 95th percentage of EEM and lumen.
In scenario of real-time semantic segmentation, frame per second (FPS) is also considered.

\begin{table}[ht]
\caption{Average performance in validation set}
\begin{center}
\begin{tabular}{c c c c c c}
\hline
\multicolumn{6}{c}{\textbf{Lumen}}\\
\hline
Method & Params & FPS & DSC & IoU & HD95 \\
\hline
PE-UNetv1$^{\mathrm{a}}$ & 40863K & 205 & 0.953 & 0.913 & 0.198 \\
PE-UNetv2$^{\mathrm{b}}$ & 40871K & 187 & 0.952 & 0.912 & 0.193 \\
CoodFCN & \textbf{475K} & 222 & 0.937 & 0.885 & 0.292 \\
MFAUNet & 646166K & 98 & 0.947 & 0.903 & 0.204 \\
IVUSNetv1 & 16952K & 179 & 0.914 & 0.846 & 0.610 \\
IVUSNetv2 & 39175K & 69 & 0.931 & 0.877 & 0.329 \\
DMUNet-D$^{\mathrm{c}}$ & 7489K & \textbf{223} & 0.934 & 0.880 & 0.306 \\
DMUNet-T$^{\mathrm{d}}$ & 7489K & \textbf{223} & 0.940 & 0.890 & 0.347 \\
DMUNet-F$^{\mathrm{e}}$ & 7489K & \textbf{223} & 0.941 & 0.892 & 0.280 \\
DLv3p-R$^{\mathrm{f}}$ & 57949K & 65 & 0.953 & 0.913 & 0.183 \\
DLv3p-X$^{\mathrm{g}}$ & 53419K & 60 & \textbf{0.957} & \textbf{0.919} & \textbf{0.177} \\
\textbf{CSDN} & 1706K & 151 & 0.953 & 0.913 & 0.189 \\
\hline
\hline
\multicolumn{6}{c}{\textbf{EEM}}\\
\hline
Method & Params & FPS & DSC & IoU & HD95 \\
\hline
PE-UNetv1$^{\mathrm{a}}$ & 40863K & 205 & 0.960 & 0.925 & 0.261 \\
PE-UNetv2$^{\mathrm{b}}$ & 40871K & 187 & 0.962 & 0.928 & 0.236 \\
CoodFCN & \textbf{475K} & 222 & 0.945 & 0.899 & 0.515 \\
MFAUNet & 646166K & 98 & 0.963 & 0.930 & 0.246 \\
IVUSNetv1 & 16952K & 179 & 0.888 & 0.803 & 1.243 \\
IVUSNetv2 & 39175K & 69 & 0.951 & 0.911 & 0.326 \\
DMUNet-D$^{\mathrm{c}}$ & 7489K & \textbf{223} & 0.945 & 0.899 & 0.550 \\
DMUNet-T$^{\mathrm{d}}$ & 7489K & \textbf{223} & 0.937 & 0.885 & 0.577 \\
DMUNet-F$^{\mathrm{e}}$ & 7489K & \textbf{223} & 0.940 & 0.890 & 0.582 \\
DLv3p-R$^{\mathrm{f}}$ & 57949K & 65 & 0.962 & 0.929 & 0.224 \\
DLv3p-X$^{\mathrm{g}}$ & 53419K & 60 & \textbf{0.964} & \textbf{0.932} & \textbf{0.213} \\
\textbf{CSDN} & 1706K & 151 & 0.960 & 0.925 & 0.246 \\
\hline
\multicolumn{6}{l}{$^{\mathrm{a,b}}$Pretrained VGG16-Encoder without BN or with BN.} \\
\multicolumn{6}{l}{$^{\mathrm{c,d,e}}$Loss is Dice, Tversky and Focal loss.} \\
\multicolumn{6}{l}{$^{\mathrm{f,g}}$Backbone is ResNet-101 or Xception.} \\
\end{tabular}
\label{tab1}
\end{center}
\end{table}

\subsection{Implementation details}
\label{ssec:sub3.3}
In all experiments, batch size in training is 16.
The size of input images is 900 $\times$ 900 pixels.
The Adam optimizer is used with an initial learning rate of $1 \times 10^{-3}$ and with a weight decay of $1 \times 10^{-4}$.
The learning rate schedule is used with a step size of 100, and learning rate is reduced by half in each step.
The total epoch we set is 300.
The models are implemented based on the PyTorch, and trained with a NVIDIA GeFore GTX 3090 GPU.
A hybrid loss combining Focal loss and Dice loss is used.
It is defined as: $$L = L_{Focal} + L_{Dice}$$ For data augmentation, a combination of the following methods is used in training:
translation, rotation, scaling, shearing, left-right and up-down flipping.
Because the channel size of input images is 3, channel swapping is also used and centeral frame is keep fixed.
Deep supervision mechanism is further used to boost segmentation performance.

\begin{figure*}[htb]
\begin{center}
\includegraphics[width=0.95\textwidth]{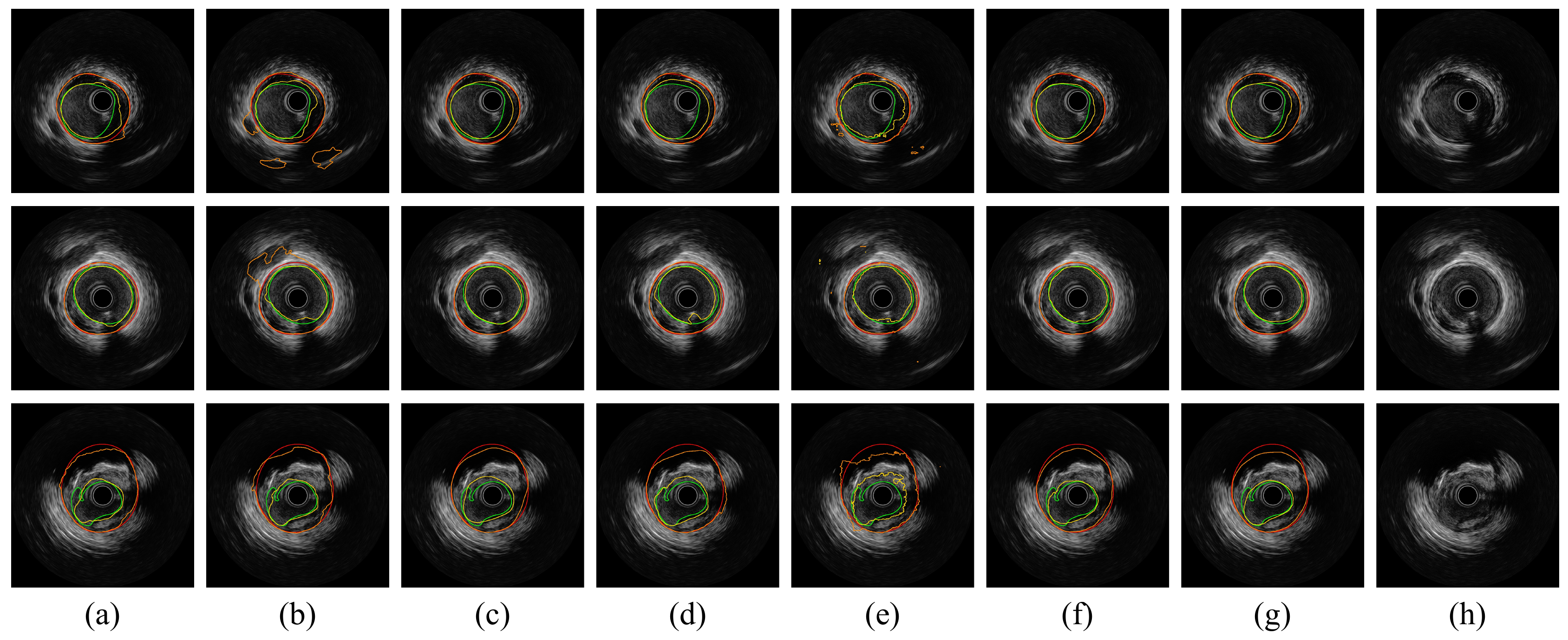}
\caption{Segmentation results from different methods. (a) PE-UNetv1 (b) CoordFCN (c) MFAUNet (d) IVUSNetv2 (e) DMUNet-F (f) DLv3p-X (g) Our CSDN (h) Input. Red contour is ground truth of EEM, and green is lumen. Orange contour is prediction of EEM, and gold is lumen.}
\label{fig2}
\end{center}
\end{figure*}

\subsection{Quantitative results}
\label{ssec:sub3.4}
Tab.~\ref{tab1} reports the quantitative results and parameter size for U-net with pretrained VGG16 encoder (PE-UNet series), CoordFCN, MFAUNet, IVUS-Netv1, IVUS-Netv2, double modified U-net with three different losses, DeepLabv3+ with different backbones and our CSDN.
Comparing to CoordFCN with 475K learnable parameters, CSDN imporves 1.6\% in DSC, 3.4\% in IoU, and 0.103 mm in HD95, for lumen segmentation, with only small increase in parameter size. For EEM segmentation, CSDN imporves 1.5\% in DSC, 2.6\% in IoU, and 0.269 mm in HD95.
Comparing to PE-UNet series, CSDN is six times less than in parameter size, but has similar or equal segmentation performance.
Although CSDN's FPS performance is less than PE-UNet series, the GPU and CPU memory usage of PE-UNet series is large than CSDN's.
It's impractical to directly deploy PE-UNet series in GPU with small memory, e.g., GTX 1050ti with 4GB.
Although MFAUNet is slightly worse than CSDN, the parameter size of MFAUNet is huge.
The size of model weights is about 2.6 GB.
It's impractical to deploy it into medical devices with small main memory.
Comparing to IVUS-Net series and DMUNet series, our CSDN surpasses in term of parameter size, region-based and border-based metrics with a large margin.
DeepLabv3+ is a semantic segmentation model for general-purpose image segmentation with large latence.
Although it processes image in real-time with GTX 3090, it's very slow or even impractical with GTX 1050ti.
Our proposed CSDN is slightly worse than DeepLabv3+, however, it's easy to deploy it into different GPU and it's smooth to segment input high-definition images in real-time.

\subsection{Qualitative results}
\label{ssec:sub3.5}
We visualize manual and predictive results for both lumen and EEM cases in Fig.~\ref{fig2}. CSDN's prediction tends to be more close to boundary of lumen and EEM, while other methods except strong but slow DeepLabv3+ with Xception as the backbone and MFAUNet with huge parameters have cases with small disconnected prediction regions or uneven contours. To demonstate the challenging task of IVUS image segmentation, the last row of Fig.~\ref{fig2} shows one of the most challenging cases where CSDN's prediction can be a reasonable segmentation from non-expert perspective. Shadow artifacts and peculiar bloodstream not only oblige learners to have capacity of modeling the global dependence, but also expect these algorithms to focus on local details. CSDN with dual streams is suitable for handling with the above dilemma.

\section{Conclusions}
\label{sec:conclusions}
We have presented CSDN, a two-stream framework for efficient and real-time segmentation of high-definition IVUS images.
It combines shallow and deep networks.
Treating low-level details and high-level semantics information separately enables learning a model to achieve high accuracy and high efficiency for accurate real-time segmentation.
In the future, more experiments can be carried out on CSDN.


 


\section{Compliance with ethical standards}
\label{sec:ethics}
Informed consent was obtained from all individual participants involved in the study.

\section{Acknowledgments}
\label{sec:acknowledgments}
This work is supported, in part, by the National Natural Science Foundation of China (Nos. 61771233).


\bibliographystyle{IEEEbib}
\bibliography{refs}

\end{document}